\documentclass[preprint,11pt]{elsarticle}

\usepackage{lineno,hyperref}
\modulolinenumbers[5]

\usepackage{amsfonts,bbm,amssymb}
\usepackage{dsfont}
\usepackage{txfonts}
\usepackage{hyperref}
\usepackage[toc,page]{appendix}

\def\bra#1{\mathinner{\langle{#1}|}}
\def\ket#1{\mathinner{|{#1}\rangle}}
\def\expect#1{\langle#1\rangle}
\def\ul#1{\underline{#1}}
\def\ol#1{\overline{#1}}

\def\bb#1{\mathbf{#1}}

\def\vmbb#1{\varmathbb{#1}}

\newcommand{\bbra}[1]{\langle\!\langle{#1}|}
\newcommand{\kket}[1]{|{#1}\rangle\!\rangle}

\newcommand{\ave}[1]{{\langle #1\rangle}}
\newcommand{\rvac}{\ket{\rm vac}}
\newcommand{\rrvac}{\kket{\rm vac}}
\newcommand{\lvac}{\bra{\rm vac}}
\newcommand{\llvac}{\bbra{\rm vac}}
\newcommand{\ra}{\rightarrow}
\newcommand{\ua}{\uparrow}
\newcommand{\da}{\downarrow}
\newcommand{\ii}{ {\rm i} }
\newcommand{\dd}{ {\rm d} }
\newcommand{\ZZ}{\mathbb{Z}}
\newcommand{\RaR}{\mathbb{R}}
\newcommand{\CC}{\mathbb{C}}

\newcommand{\z}{{\rm z}}

\newcommand{\PP}{{\hat{\cal P}}}
\newcommand{\LL}{{\hat{\cal L}}}

\newcommand{\DD}{{\hat{\cal D}}}

\newcommand{\eref}[1]{(\ref{#1})}
\def\tr{{{\rm tr}}}

\def\End{{\,{\rm End}\,}}
\def\one{\mathbbm{1}}

\usepackage{fancyhdr}
\journal{Nuclear Physics B}

\begin{document}

\begin{frontmatter}

\title{Exact steady state manifold of a boundary driven spin-$1$ Lai--Sutherland chain}
\author{Enej Ilievski and Toma\v{z} Prosen}

\address{Department of Physics, Faculty of Mathematics and Physics, University of Ljubljana,
Jadranska 19, SI-1000 Ljubljana, Slovenia}

\begin{abstract}
We present an explicit construction of a family of steady state density matrices for an open 
integrable spin-$1$ chain with bilinear and biquadratic interactions, also known as the Lai--Sutherland model, driven far from equilibrium by means of
two oppositely polarizing Markovian dissipation channels localized at the boundary. 
The steady state solution exhibits $n+1$ fold degeneracy, for a chain of length $n$, due to existence of (strong)
Liouvillian $U(1)$ symmetry. The latter can be exploited to introduce a chemical potential and define a grand canonical nonequilibrium steady state ensemble.
The matrix product form of the solution entails an infinitely-dimensional representation of
a non-trivial Lie algebra (semidirect product of $\mathfrak{sl}_2$ and a non-nilpotent radical) and hints to a novel Yang-Baxter integrability structure.
\end{abstract}
\end{frontmatter}

%\linenumbers

\section{Introduction}
\label{Intro}

Nonequilibrium  transport problem in extended low-dimensional (say one-dim\-ensional, 1D) quantum systems is an important current topic in statistical mechanics with possible links to experiments in condensed matter systems \cite{spinchains,fabian}.
Among the most important open issues are (i) classification or identification of possible transport behaviors, ranging from ballistic, via diffusive (normal or anomalous), to insulating, and understanding their microscopic mechanisms \cite{affleck,marko},
and (ii) developing nonequilibrium quantum thermodynamics \cite{gemmer} and a theory of nonequilibrium quantum phase transitions (see e.g. \cite{prl108,zoller}).

A convenient setup for studying far from equilibrium {\em relaxation dynamics}  or {\em steady state} situations which support macroscopic currents of charge/particles, magnetization, or energy/heat,
is to couple a 1D system of strongly interacting quantum particles to incoherent forcing governed by
two reservoirs attached at each end of the particle chain and assign them different effective thermodynamic potentials. This can be achieved e.g.
by choosing simple Markovian dissipation channels which operate as quantum jumps, i.e. pump-in or absorb-out the elementary excitations at the surface (boundary of the chain). The rest of the
system is chosen to be unaffected by the dissipation and hence evolves according to fully-coherent unitary evolution. For a derivation and physical justification of such an approach, see Refs.\cite{gemmer,gemmer2}.

With the hope of being able to take advantage of their rich and elegant mathematical content, one addresses {\em integrable} systems with {\em local} interactions first.
In this light one obtains a toy model to study dissipative integrable theory with surface `non-unitary sources'. This model could be in some
sense also regarded as a quantum analogue of classical stochastic exclusion processes \cite{derrida,schutz,blythe}.
Focusing initially on the steady states alone, the aim is to be able to isolate regimes where complexity of the steady state density operator is drastically reduced,
opening a possibility of finding an efficient and exact representation in terms of the matrix product state.
Ever since the first solutions in this direction have been presented, addressing quasi-free theory \cite{njp} and the paradigmatic strongly interacting case of the (anisotropic) Heisenberg
($XXZ$) spin-$1/2$ chain \cite{prl11a,prl11b}, the quest for new integrable out-of-equilibrium scenarios continues \cite{kps,prl14}, with some recent attempts \cite{pip13,iz,prl13} of putting 
these searches under the common roof of theory of integrable quantum systems \cite{sutherland_book,korepin}.

It has been argued \cite{kps,iz} that explicit steady state solutions of boundary driven Liouvillian (Lindbladian) flows pertaining to certain integrable models arise as a consequence of the underlying quantum group symmetry of the model.
The latter provides a prerequisite condition for the solution in the bulk which needs to be fine-tuned with the form of the quantum noise process applied to the system's boundaries.
Two principal insights have been made, namely (i) to re-write the matrix product representation of solution in terms of monodromy matrices
with Lax operators arising from solutions of the universal Yang-Baxter equation associated to a symmetry algebra of an interaction, and (ii) to allow for non-unitary irreducible representations over infinitely-dimensional vector spaces \cite{derkachev}.
In the prototype case of the Heisenberg spin-$1/2$ model it actually appears that fundamental (local) building blocks that generate the solution
inherit symmetry from the interaction, despite that the latter is finally broken at the level of Liouvillian flow and density operator.
Below we demonstrate however, how central objects of our construction could also admit a (non-trivial) continuous symmetry which does {\em not} respect that of
an integrable bulk interactions.

To this end, we consider an integrable {\em SU(3)}-invariant spin-$1$ chain, commonly referred to as the \textit{Lai--Sutherland model} \cite{lai,sutherland}\footnote{Despite its commonly known name, the model has
been discussed even a few years earlier by Uimin \cite{uimin}.},
and employ a pair of Lindblad jump operators which couple only two extreme levels at the chains end.
The intermediate level, which can be viewed as a hole particle, is thus protected from the environment and its number
is preserved throughout the (dissipative) evolution.
Henceforth, such Lindbladian flow is \textit{reducible} to a (thermodynamically) infinite number of sectors corresponding to subspaces with
fixed `hole doping'. This allows for a possibility of constructing a {\em grand-canonical steady state ensemble},
with chemical potential being an additional parameter which controls the average number of holes (or the filling factor).

The paper is organized as follows. In section \ref{model} we introduce the open three-state Lai--Sutherland model and specify suitable `integrable' boundary dissipative processes in the framework of Markovian (Lindblad)
master equations. In section \ref{solution} we rigorously construct the solution of the steady state in terms of an infinite rank matrix product ansatz.
In section \ref{remarks} we discuss several important physical properties of the solution: we introduce the grand-canonical nonequilibrium steady state ensemble (subsection \ref{grand_canonical}), 
describe a formal computation of local physical observables (subsection \ref{observables}), discuss graph-theoretic interpretation of the solution in terms of sums over walks (subsection \ref{walks}), characterize the symmetries
(subsections \ref{lie_algebra}, \ref{symmetry_lax}, \ref{symmetries}) and discuss possible connection to quantum inverse scattering method  (\ref{transfer_matrix}). Finally, we conclude in section \ref{conclusions}.

\section{Open Lai--Sutherland model far from equilibrium}
\label{model}

Consider a finite chain of $n$ sites and let $\mathfrak{H}_1\cong \CC^{3}$ be a local quantum (`physical') space associated to each spin-site $x \in\{1,2,\ldots,n\}$. The entire $3^n$ dimensional
many-body quantum space $\mathfrak{H}_{s}$ is constructed as $n$-fold tensor product of local spaces,
$\mathfrak{H}_{\rm s}=\mathfrak{H}_1^{\otimes n}$.
Using the Weyl matrix basis $\{ e^{ij} = \ket{i}\!\bra{j}; i,j=1,2,3\}$ of $\End(\mathfrak{H}_1) = \frak{gl}_3$, we define a full set of local generators of the matrix algebra $\mathfrak{F}=\End(\mathfrak{H}_{\rm s})$ as
\begin{equation}
e^{ij}_x = \one_3^{\otimes (x-1)} \otimes e^{ij} \otimes \one_3^{\otimes (n-x)},
\end{equation}
$\one_d$ defining a $d-$dimensional unit matrix, satisfying the Lie algebra relations
\begin{equation}
[e_{x}^{ij},e_{x'}^{kl}]=(\delta_{j\,k}e_{x}^{i\,l}-\delta_{i\,l}e_{x}^{kj})\delta_{x,x'}.
\end{equation}
The spin-$1$ Lai--Sutherland model \cite{lai,sutherland} for a chain of $n$ sites is given by the Hamiltonian $H\in \mathfrak{F}$,
\begin{equation}
H=\sum_{x=1}^{n-1}h_{x,x+1}, \quad h_{x,x+1}= \vec{s}_{x}\cdot \vec{s}_{x+1}+(\vec{s}_{x}\cdot \vec{s}_{x+1})^{2} - \one,
\label{Lai--Sutherland}
\end{equation}
where $\vec{s}_x=(s^{1}_x,s^{2}_x,s^{3}_x)$, with 
\begin{equation}
s_x^1 = \frac{1}{\sqrt{2}}(e_x^{12}+e_x^{21}+e_x^{23}+e_x^{32}),\; s^2_x = \frac{\ii}{\sqrt{2}}(e_x^{21}-e_x^{12}+e_x^{32}-e_x^{23}),\; s^3_x=e_x^{11}-e_x^{33},
\end{equation}
form independent spin-$1$ variables (local $s=1$ representations of $\mathfrak{su}_2$) satisfying
\begin{equation}
[s_{x}^{i},s_{x'}^{j}]=\ii\sum_k \epsilon_{ijk}s_{x}^{k}\delta_{x,x'}.
\end{equation}
Straightforward inspection shows that the local Hamiltonian $h_{x,x+1}$ -- the interaction -- is in fact just the permutation operator between neighboring sites 
\begin{equation}
h_{x,x+1} = \sum_{i,j=1}^3 \one_3^{\otimes(x-1)}\otimes \ket{i,j}\bra{j,i}\otimes \one_3^{\otimes (n-x-1)}= \sum_{i,j=1}^3 e^{ij}_x e^{j\,i}_{x+1}.
\end{equation}
The local Hilbert state basis is therefore given by a triple of states $\ket{1}\equiv \ket{\uparrow},\ket{2}\equiv \ket{0},\ket{3}\equiv \ket{\downarrow}$,
which can be interpreted as three different particle species; respectively, as {\em spin-up} particles,  {\em holes}, and {\em spin-down} particles.
The model then becomes equivalent to the so-called {\em supersymmetric t-J model} \cite{tJ}.

Lai--Sutherland chain is a multi-component quantum model and we may associate with it a skew-symmetric tensor of particle currents, with two-site density
\begin{equation}
J^{ij} = \ii (e^{ij}\otimes e^{j\,i} - e^{j\,i}\otimes e^{ij}),\quad J^{ij}_x = \one^{\otimes (x-1)}_3 \otimes J^{ij} \otimes \one^{\otimes(n-1-x)}_3 = -J^{j\,i}_x,
\end{equation}
which, by construction, satisfies the following continuity equation
\begin{equation}
\frac{\dd}{\dd t} (e^{i\,i}_x - e^{jj}_x) = \ii [H,e^{i\,i}_x - e^{jj}_x] = J^{ij}_{x-1,x} - J^{ij}_{x,x+1}.
\label{continuity_equation}
\end{equation}
$J^{ij}$ can be considered as a partial current of the particle of species $i$ into particles of species $j$.
The total current of particles of species $i$,
\begin{equation}
J^i = \sum_{j=1}^3 J^{ij},
\label{totalJ}
\end{equation}
then fulfills the continuity equation
\begin{equation}
\frac{\dd}{\dd t} e^{i\,i}_x = J^i_{x-1,x}  - J^i_{x,x+1},
\end{equation}
where $e^{i\,i}_x$ can be considered as the operator of particle density of species $i$.

We shall now open the Lai--Sutherland chain and couple it to the environment via Markovian processes which act only on local quantum spin spaces at the boundary, i.e., at $x=1$ and $x=n$. The many-body density operator
$\rho_t$, $t\in\RaR^+$, considered as an element of $\mathfrak{F}$ which may be here considered as a Liouville vector space of operators, then evolves according to Liouvillian semigroup
\begin{equation}
\rho_t(\varepsilon)=\exp{(t \LL)}\rho_0,\qquad \LL=\LL_{0}+\varepsilon \DD,
\label{Lindblad_equation}
\end{equation}
with time-independent generator -- the Liouvillian $\LL\in \End{(\mathfrak{F})}$ being split into non-dissipative part $\LL_{0}(\rho)\equiv -\ii [H,\rho]$ governing {\em unitary}
Liouville--von Neumann evolution, and a dissipator $\DD\in \End{(\mathfrak{F})}$ describing the incoherent, dissipative (non-unitary) processes of overall strength $\varepsilon$. The latter is given in terms
of a set of jump operators $\{A_{\alpha} \in \mathfrak{F}\}$ and takes a general canonical Lindblad \cite{lindblad,gorini} form
\begin{equation}
\DD \rho=\sum_{\alpha}\DD_{A_\alpha}(\rho),\quad{\rm where}\quad
\DD_A(\rho) :=  2A \rho A^{\dagger} -\{A^{\dagger}A,\rho\}.
\label{D}
\end{equation}
In particular, we install a single {\em local} jump operator at each end of the chain:
\begin{equation}
A_{1}=e_{1}^{13} = \frac{1}{2} (s^+_1)^2,\quad A_{2}=e_{n}^{31} = \frac{1}{2} (s^-_n)^2,\;\;{\rm where}\;\;
s^\pm_x :=  s^1_x \pm \ii s^2_x.
\label{driving}
\end{equation} 
Two dissipation channels, interpreted as the left and right magnetization bath, perform the processes 
$\ket{\uparrow}\to\ket{\downarrow}$ and $\ket{\downarrow}\to\ket{\uparrow}$, respectively, with the rates $\varepsilon$. Both processes keep the hole state $\ket{0}$ unaffected.
Since also the bulk dynamics generated by $\LL_0$ conserves the number of particles of each species, it follows that the whole Liouvillian dynamics (master equation) preserves the number of holes.
More precisely, defining the hole-number operator $N_0 \in \mathfrak{F}$ as
\begin{equation}
N_0 \ket{i_1,i_2,\ldots,i_n} = \left(\sum_{x=1}^n \delta_{i_x,2}\right) \ket{i_1,i_2,\ldots,i_n},
\end{equation}
we have that the set of all, Hamiltonian and jump operators, commute with $N_0$
\begin{equation}
[H,N_0] = 0,\quad [A_{1,2},N_0] = 0,
\end{equation}
which implies that $N_0$ generates a {\em strong} \cite{buca} $U(1)$ symmetry of the Liouvillian flow (\ref{Lindblad_equation}).
$N_0$ foliates the physical space into $n+1$ orthogonal eigenspaces, $\mathfrak{H}_{\rm s} = \bigoplus_{\nu=0}^n \mathfrak{H}^{(\nu)}_{\rm s}$,
$N_0 \mathfrak{H}^{(\nu)}_{\rm s} = \nu \mathfrak{H}^{(\nu)}_{\rm s}$. The theorem A.1 of Ref.~\cite{buca} then guarantees that the full Lindblad dynamics (\ref{Lindblad_equation})
is closed on $\mathfrak{F}^{(\nu)}=\End(\mathfrak{H}^{(\nu)}_{\rm s})$, $\LL^{(\nu)} = \LL|_{\mathfrak{F}^{(\nu)}}$,
and that a {\em fixed point} $\rho^{(\nu)}_\infty = \lim_{t\to \infty} \exp(t \LL^{(\nu)}) \rho^{(\nu)}_0$ -- {\em nonequilibrium steady state} (NESS) -- exists for each symmetry subspace flow
\footnote{Note that Thm. A.1 of \cite{buca} guarantees that dynamics (\ref{Lindblad_equation}) is closed inside {\em non-diagonal} spaces ${\rm Lin}(\mathfrak{H}^{(\nu)},\mathfrak{H}^{(\nu')})$, $\nu\neq\nu'$, as well, but these may or may not \cite{albert} (based on computer experiments we conjecture that they {\em do not})
support Liouvillian fixed points and shall not be discussed in this paper.},
\begin{equation}
\LL^{(\nu)} \rho^{(\nu)}_\infty = -\ii [H, \rho^{(\nu)}_{\infty}] + \varepsilon \DD(\rho^{(\nu)}_{\infty}) = 0. \label{fixed_point}
\end{equation}
The theorem by Evans \cite{evans} can then be used to show uniqueness of NESS $\rho^{(\nu)}_\infty$ for each fixed $\nu$.
In the next section we shall outline a simple algebraic procedure for actual explicit construction of density operators $\rho^{(\nu)}_\infty$.

\section{Matrix product solution}
\label{solution}

Let $\PP^{(\nu)}\in\End(\mathfrak{F})$ be an orthogonal projector to $\mathfrak{F}^{(\nu)}$. We define a {\em universal} density matrix of NESS as a direct sum of non-trivial solutions of (\ref{fixed_point}) for all $\nu$,
\begin{equation}
\rho_\infty = \sum_{\nu=0}^{n} \rho^{(\nu)}_\infty,\quad{\rm with}\quad\rho^{(\nu)}_\infty= \PP^{(\nu)}\rho_\infty \neq 0,
\label{universal}
\end{equation}
being solution of the fixed point equation (\ref{fixed_point}) as well. The state $\rho_\infty$ shall be sought for in terms of Cholesky factorization
(in analogy to previous solutions of $XXZ$ \cite{prl11b} and Hubbard \cite{prl14} models)
\begin{equation}
\rho_{\infty}(\varepsilon)=S_n(\varepsilon)S^{\dagger}_n(\varepsilon),
\label{factorization}
\end{equation}
where $S_{n}(\varepsilon)\in \End(\mathfrak{H}_{\rm s})$ is some yet unknown operator which is represented by an upper triangular matrix in the computational basis $\ket{i_1,\ldots,i_n}$.
Introducing an auxiliary Hilbert space $\mathfrak{H}_{\rm a}$ -- separable, but of infinite dimensionality as will become clear later -- 
we define the monodromy operator
$\bb{M}(\varepsilon)\in \End{(\mathfrak{H}_{\rm s}\otimes \mathfrak{H}_{\rm a})}$ as a spatially-ordered product of some local Lax operators
\footnote{A suggestive name {\em Lax operator} should hint on the relation to the zero-curvature condition which shall be established later.}
$\bb{L}_{x}(\varepsilon)\in \End{(\mathfrak{H}_{\rm s}\otimes \mathfrak{H}_{\rm a})}$,
\begin{equation}
\bb{M}(\varepsilon) =\bb{L}_{1}(\varepsilon)\bb{L}_{2}(\varepsilon)\cdots \bb{L}_{n}(\varepsilon).
\end{equation}
Throughout the paper, the upright-boldface notation designates object which are not scalars in auxiliary space $\mathfrak{H}_{\rm a}$.  Index free Lax operator can be defined as $\bb{L}(\varepsilon) \in \End(\mathfrak{H}_1\otimes\mathfrak{H}_{\rm a})$ so that one writes 
$\bb{L}_{x}(\varepsilon)=\one_{3}^{\otimes (x-1)} \otimes \bb{L}(\varepsilon)\otimes \one_{3}^{\otimes (n-x)}$. 
Furthermore, we define the components of Lax matrix $\bb{L}^{ij}(\varepsilon) \in  \End(\mathfrak{H}_{\rm a})$, such that
\begin{equation}
\bb{L}_{x}(\varepsilon)=\sum_{i,j=1}^{3}e_{x}^{ij}\otimes \bb{L}^{ij}(\varepsilon),\qquad
\bb{L}(\varepsilon)=\sum_{i,j=1}^{3}e^{ij}\otimes \bb{L}^{ij}(\varepsilon).
\label{L_operator}
\end{equation}
We further assume existence of a special state $\ket{\rm vac}\in\mathfrak{H}_{\rm a}$, such that Cholesky factor writes as the auxiliary expectation value of monodromy operator, or equivalently, as a {\em matrix product operator} (MPO)
\begin{equation}
S_n(\varepsilon) = \bra{\rm vac}\bb{M}(\varepsilon)\ket{\rm vac} = \sum_{i_1,j_1\ldots i_n,j_n} \bra{\rm vac}\bb{L}^{i_1 j_1}\cdots \bb{L}^{i_n j_n}\ket{\rm vac} e^{i_1 j_1}\otimes \cdots \otimes e^{i_n j_n}.
\label{MPO}
\end{equation}
Fixing an arbitrary, fixed orthonormal basis $\{\ket{\psi_k}\}$ of $\mathfrak{H}_{\rm a}$ we define the conjugate Lax matrices $\ol{\bb{L}}(\varepsilon)$ by 
$\bra{\psi_k}\ol{\bb{L}}^{ij}(\varepsilon)\ket{\psi_l} := \ol{\bra{\psi_k}\bb{L}^{ij}(\varepsilon)\ket{\psi_l}}$.
For notational convenience we denote the second copy of auxiliary space carrying conjugate representation of $\ol{\bb{L}}^{ij}$ as $\ol{\mathfrak{H}}_{\rm a}$.
One can then write MPO formulation of NESS density operator $\rho_{\infty}$ directly, by introducing {\em two-leg Lax matrices} $\vmbb{L}^{ij}(\varepsilon)\in\End(\mathfrak{H}_{\rm a}\otimes \ol{\mathfrak{H}}_{\rm a})$,
and $\vmbb{L}_x(\varepsilon) \in \End(\mathfrak{H}_{\rm s}\otimes \mathfrak{H}_{\rm a}\otimes \ol{\mathfrak{H}}_{\rm a})$ as
\begin{equation}
\vmbb{L}^{ij}(\varepsilon) = \sum_{k} \bb{L}^{ik}(\varepsilon)\otimes \ol{\bb{L}}^{jk}(\varepsilon),\quad \vmbb{L}_x(\varepsilon) = \sum_{i,j} e^{ij}_x \otimes \vmbb{L}^{ij}(\varepsilon),
\label{double_L}
\end{equation}
namely
\begin{equation}
\rho_{\infty}(\varepsilon)=\llvac\vmbb{M}(\varepsilon)\rrvac.
\label{rhoMM}
\end{equation}
Note the transposition in the quantum space of the conjugated factor of \eref{double_L}.
Here a two-leg monodromy operator 
\begin{equation}
\vmbb{M}(\varepsilon) = \vmbb{L}_1(\varepsilon)\cdots  \vmbb{L}_n(\varepsilon) \in \End(\mathfrak{H}_{\rm s}\otimes\mathfrak{H}_{\rm a}\otimes \ol{\mathfrak{H}}_{\rm a}),
\label{MM}
\end{equation}
and a product of a pair of vacua $\llvac=\lvac \otimes \lvac$, $\rrvac=\rvac \otimes \rvac$ have been introduced, so that (\ref{rhoMM}) is merely a formal rewriting of (\ref{factorization}).
These definitions become particularly handy when we consider evaluation of expectation values of local observables with respect to NESS $\rho_{\infty}(\varepsilon)$.

Let $\eta:=\ii \varepsilon$ be a complex-rotated coupling parameter and let us (for convenience) relabel the quantum space matrix elements of the $\bb{L}$-operator as
\begin{eqnarray}
\label{relabeling}
\bb{L}=\pmatrix{
\bb{l}^{\ua} & \bb{t}^{+} & \bb{v}^{+} \cr
\bb{t}^{-} & \bb{l}^{0} & \bb{u}^{+} \cr
\bb{v}^{-} & \bb{u}^{-} & \bb{l}^{\da}
}.
\end{eqnarray}
The key results of this paper are the following:
\newtheorem{thm}{Theorem}
\begin{thm}
Suppose that 9 matrix elements $\{\bb{L}^{ij}\}$ generate the Lie algebra $\mathfrak{g}$ defined by commutation relations,
\begin{eqnarray}
\label{algebra}
& [\bb{u}^{+},\bb{t}^{\pm}]=[\bb{u}^{-},\bb{t}^{\pm}]=[\bb{u}^{\pm},\bb{v}^{\pm}]=[\bb{t}^{\pm},\bb{v}^{\pm}]=0,\cr
& [\bb{l}^{\ua},\bb{u}^{\pm}]=[\bb{l}^{\da},\bb{t}^{\pm}]=[\bb{l}^{\ua},\bb{l}^{\da}]=0,\cr
& [\bb{l}^{\ua},\bb{t}^{\pm}]=\mp \eta \bb{t}^{\pm},\quad [\bb{l}^{\da},\bb{u}^{\pm}]=\mp \eta \bb{u}^{\pm},\cr
&\!\!\![\bb{u}^{\pm},\bb{v}^{\mp}]=\pm \eta \bb{t}^{\mp},\,\quad [\bb{t}^{\pm},\bb{v}^{\mp}]=\pm \eta \bb{u}^{\mp}, \cr
&  [\bb{l}^{\ua},\bb{v}^{\pm}]=[\bb{l}^{\da},\bb{v}^{\pm}]=\mp \eta \bb{v}^{\pm},\quad [\bb{v}^{+},\bb{v}^{-}]=\eta(\bb{l}^{\ua}+\bb{l}^{\da}),\cr
& [\bb{t}^{+},\bb{t}^{-}]=[\bb{u}^{+},\bb{u}^{-}]=\eta \bb{l}^{0},\cr
& [\bb{l}^{\ua,\da},\bb{l}^{0}] = [\bb{u}^\pm,\bb{l}^{0}] = [\bb{v}^\pm,\bb{l}^{0}] = [\bb{t}^\pm,\bb{l}^{0}] = 0,
\end{eqnarray}
with a representation over the Hilbert space $\mathfrak{H}_{\rm a}$ satisfying the following conditions
\begin{eqnarray}
\label{boundary_conditions}
&\bb{l}^{\ua}\rvac = \bb{l}^{0}\rvac = \bb{l}^{\da}\rvac = \rvac,\cr
&\lvac \bb{l}^{\ua} = \lvac \bb{l}^{0} = \lvac \bb{l}^{\da} = \lvac,\cr
&\bb{t}^+ \rvac = \bb{u}^+ \rvac = \bb{v}^+ \rvac = 0,\cr
&\lvac \bb{t}^- = \lvac \bb{u}^- = \lvac\bb{v}^- = 0.
\end{eqnarray}
Then, the universal solution (\ref{universal}) to NESS fixed point condition \eref{fixed_point} is given via Cholesky factorization \eref{factorization} with explicit MPO expression \eref{MPO} for $S_n(\varepsilon)$ with $\eta=\ii\varepsilon$.
\end{thm}

\begin{thm}
A possible irreducible explicit representation of Lie algebra $\mathfrak{g}$ \eref{algebra} satisfying \eref{boundary_conditions} is given as
\begin{eqnarray}
& \bb{t}^{+}=\bb{b}_{\ua},\quad \bb{t}^{-}=\eta \bb{b}^\dagger_{\ua}, \cr
& \bb{u}^{+}=\eta \bb{b}_{\da},\quad \bb{u}^{-}=\bb{b}^\dagger_{\da},\cr
& \bb{v}^{+}=\eta(\bb{b}_{\ua}\bb{b}_{\da}+\bb{s}^{+}),\quad \bb{v}^{-}=\eta(\bb{b}^\dagger_{\ua}\bb{b}^\dagger_{\da}-\bb{s}^{-}),\cr
& \bb{l}^{\ua,\da}=\eta\left(\bb{b}^\dagger_{\ua,\da}\bb{b}_{\ua,\da}+\frac{1}{2}-\bb{s}^{\rm z}\right),\quad \bb{l}^{0}=\one,
\label{algebra_generators}
\end{eqnarray}
in terms of three auxiliary degrees of freedom with a three dimensional lattice $\{\ket{j,k,l},\; j,k,l\in\ZZ^+\}$ forming a basis of $\mathfrak{H}_{\rm a}$, namely, two bosonic modes $\bb{b}_{\ua,\da}$
\begin{eqnarray}
\bb{b}^\dagger_{\ua}\ket{j,k,l}= \sqrt{j+1}\ket{j+1,k,l},\quad \bb{b}_{\ua}\ket{j,k,l}=\sqrt{j}\ket{j-1,k,l},\cr
\bb{b}^\dagger_{\da}\ket{j,k,l}= \sqrt{k+1}\ket{j,k+1,l},\quad \bb{b}_{\da}\ket{j,k,l}=\sqrt{k}\ket{j,k-1,l},
\label{bosons}
\end{eqnarray}
and a complex spin (Verma module of $\mathfrak{sl}_2$)
\begin{eqnarray}
\bb{s}^{+} \ket{j,k,l} &=& l \ket{j,k,l-1},\cr 
\bb{s}^-\ket{j,k,l} &=& (2p-l) \ket{j,k,l+1},\cr 
\bb{s}^{\rm z}\ket{j,k,l} &=& (p-l)\ket{j,k,l}.
\label{verma}
\end{eqnarray}
with $\rvac = \ket{0,0,0}$ being the highest-weight-state.
The complex spin parameter $p$ should be linked to dissipation parameter via
\begin{equation}
p = \frac{1}{2} - \frac{1}{\eta} = \frac{1}{2} + \frac{\ii}{\varepsilon}.
\label{pfix}
\end{equation}
\end{thm}

\newproof{pf}{Proof}
\begin{pf}
The proof of the theorems is based on verifying that the Lie algebra $\mathfrak{g}$, given by \eref{algebra}, can be equivalently defined by means of an identity over
$\End(\mathfrak{H}_{\rm s}\otimes \mathfrak{H}_{\rm a})$ in the form of \textit{local operator divergence} (LOD) condition (customary referred to as the
Sutherland equation which is equivalent to {\em zero curvature}/Lax condition),
\begin{equation}
[h_{x,x+1},\bb{L}_{x}(\varepsilon)\bb{L}_{x+1}(\varepsilon)]=B_{x}(\varepsilon)\bb{L}_{x+1}(\varepsilon)-\bb{L}_{x}(\varepsilon)B_{x+1}(\varepsilon),
\label{Sutherland}
\end{equation}
with the-so-called {\em boundary operator} $B_{x}(\varepsilon) \in \End{(\mathfrak{H}_{\rm s}\otimes \mathfrak{H}_{\rm a})}$ -- operating non-trivially only in the local quantum space
\begin{equation}
B_{x}=\eta \left(e_{x}^{33} \otimes \one_{\rm a} - e_{x}^{11}\otimes \one_{\rm a}\right)= b_x \otimes \one_{\rm a}, \quad {\rm where}\quad
b_x(\varepsilon) = -\ii\varepsilon s^{3}_x \in \mathfrak{F}.
\label{B}
\end{equation}
Identification of \eref{algebra} with LOD \eref{Sutherland} is straightforward, based solely on the permutation action of Hamiltonian density 
\begin{equation}
[h_{x,x+1},e^{ij}_x e^{kl}_{x+1}] = e^{kj}_x e^{i\,l}_{x+1} - e^{i\,l}_x e^{kj}_{x+1}.
\end{equation}
Multiplying LOD by a string $\bb{L}_1\cdots \bb{L}_{x-1}$ from the left and a string $\bb{L}_{x+2}\cdots \bb{L}_n$ from the right, summing over $x$
and taking vacuum expectation value yields the global almost conservation condition for the Cholesky factor (the so-called {\em defining relation}, analogous to similar relations in other integrable nonequilibrium models \cite{prl11b,prl14}),
\begin{equation}
[H,S_{n}(\varepsilon)]=-\ii\varepsilon\left(s^{3}\otimes S_{n-1}(\varepsilon)-S_{n-1}(\varepsilon)\otimes s^{3}\right), \quad {\rm where}\quad s^3 = e^{11}-e^{33}.
\label{defining_relation}
\end{equation}
Consequently, by expanding the unitary part of Liouvillian $\LL_{0}$,
\begin{equation}
-\LL_{0}(\rho_{\infty})\equiv \ii[H,\rho_{\infty}]=\ii[H,S_{n}]S^{\dagger}_{n}-\ii S_{n}[H,S_{n}]^{\dagger},
\end{equation}
in conjunction with \eref{defining_relation}, and employing the definition \eref{double_L}, the steady state condition \eref{fixed_point} yields
a decoupled system of \textit{boundary equations} 
\begin{eqnarray}
& \llvac\left( \DD_{A_1}(\vmbb{L}_{1}) - \ii(\vmbb{B}^{(1)}_{1}-\vmbb{B}^{(2)}_{1})\right) = 0,\nonumber\\
& \left(\DD_{A_2}(\vmbb{L}_{n}) + \ii(\vmbb{B}^{(1)}_{n}-\vmbb{B}^{(2)}_{n})\right)\rrvac = 0,
\label{boundary_system}
\end{eqnarray}
where {\em two-leg boundary operators} $\vmbb{B}^{(1)}_{x},\vmbb{B}^{(2)}_{x}\in \End(\mathfrak{H}_{\rm s}\otimes \mathfrak{H}_{\rm a}\otimes \ol{\mathfrak{H}}_{\rm a})$, reading 
\begin{equation}
\vmbb{B}^{(1)}_{x}= \sum_{i,j=1}^3 b_x e^{ij}_x \otimes \one_{\rm a}\otimes \ol{\bb{L}}^{j\,i},\qquad 
\vmbb{B}^{(2)}_{x}= \sum_{i,j=1}^3 e^{ij}_x \ol{b}_x \otimes \bb{L}^{ij} \otimes \one_{\rm a},
\end{equation}
have been defined. Note that, due to \eref{B}, $\ol{b}_x = \ii \varepsilon s^3_x = -b_x$ for $\varepsilon\in\RaR$.

The last two lines of  \eref{algebra} indicate that pairs of auxiliary operators $(\bb{t}^{+},\bb{t}^{-})$ and $(\bb{u}^{+},\bb{u}^{-})$ span the Weyl-Heisenberg algebra.
In conjunction with the highest weight conditions \eref{boundary_conditions} this fixes (uniquely, up to unitary transformations) the representation of  $(\bb{t}^{+},\bb{t}^{-})$ and $(\bb{u}^{+},\bb{u}^{-})$ to be that of a
 Fock space of two canonical bosonic (oscillator) modes, specified by creation/annihilation operators,
$[\bb{b}_{\sigma},\bb{b}^{\dagger}_{\sigma'}]=\delta_{\sigma,\sigma'}$, $[\bb{b}_{\sigma},\bb{b}_{\sigma'}]=0$, $\sigma,\sigma'\in \{\ua,\da\}$, suggesting that the auxiliary space $\mathfrak{H}_{\rm a}$ is perhaps just a two-mode boson Fock space.
While realization for all the other generators consistent with the bulk algebra $\mathfrak{g}$ is not difficult to construct (e.g. $\bb{v}^\pm$, $\bb{l}^\ua+\bb{l}^\da$ can be just the Schwinger boson representation of $\mathfrak{su}_2$ -- see 5th line of \eref{algebra}),
it turns out not to be consistent with the boundary conditions \eref{boundary_conditions}
\footnote{One can for instance compute Schmidt ranks of bipartite (symmetric) cut for exact MPO solution of $S_n$ for small systems sizes and observe that
they exceed the upper bounds implied by the conjectured two-particle Fock space for $\mathfrak{H}_{\rm a}$.}. Therefore the auxiliary space $\mathfrak{H}_{\rm a}$ has to contain (at least) one additional degree of freedom.

Ultimately, in order to fulfill \eref{boundary_system}, a straightforward calculation shows that it is enough to add a Verma module $\mathfrak{S}$ of complex spin representation \eref{verma} of $\mathfrak{sl}_2$ and consider a triple-product space $\mathfrak{H}_{\rm a}\cong \mathfrak{B}\otimes \mathfrak{B}\otimes \mathfrak{S}={\rm lsp}\{ \ket{j,k,l}; j,k,l\in\ZZ^+\}$,
and find a representation of the algebra \eref{algebra} which is compliant with conditions
\begin{eqnarray}
\label{Lax_boundary}
\bb{L}\rvac =
\pmatrix{
\rvac & 0 & 0 \cr
\eta \ket{1,0,0} & \rvac & 0 \cr
\eta(\ket{1,1,0}-\ket{0,0,1})+2\ket{0,0,1} & \ket{0,1,0} & \rvac},\\
\lvac \bb{L} =
\pmatrix{
\lvac & \bra{1,0,0} & \eta(\bra{1,1,0}+\bra{0,0,1}) \cr
0 & \lvac & \eta \bra{0,1,0} \cr
0 & 0 & \lvac},
\end{eqnarray}
with vacuum being given by the ground state $\rvac \equiv \ket{0,0,0}$.
These requirements are all satisfied by choosing representation (\ref{algebra_generators},\ref{bosons},\ref{verma}) with $p$ being fixed (\ref{pfix}) as
required by the conditions in the first two lines of (\ref{boundary_conditions}). The last two lines of (\ref{boundary_conditions}) hold due to highest-weight-property of $\rvac$. As such a representation is clearly irreducible, this concludes the proof of theorems 1 and 2.
\end{pf}

\noindent
{\sc Remark.} All MPO (\ref{MPO}) amplitues, i.e., matrix elements of the Cholesky factor of the density operator
\begin{equation}
\bra{i_1,i_2,\ldots,i_n}S_n \ket{j_1,j_2,\ldots,j_n} = \lvac \bb{L}^{i_1 j_1}\bb{L}^{i_2 j_2}\cdots \bb{L}^{i_n j_n}\rvac,
\end{equation}
are polynomials (of order not more than $n$) in $\eta=\ii\varepsilon$ with {\em integer} coefficients. This is a simple consequence of Wick theorem, or representation of Theorem 2.

\section{Discussion}
\label{remarks}

The formulae (\ref{factorization},\ref{MPO},\ref{relabeling},\ref{algebra_generators},\ref{bosons},\ref{verma},\ref{pfix}) are the main result of this paper:
They generate explicit construction of a many-body density matrix of a family of degenerate NESSes $\rho^{(\nu)}_\infty = \PP^{(\nu)}\rho_\infty$ for any number of holes $\nu\in\{0,1\ldots n\}$. The computational complexity of obtaining any locality-based information about the state $\rho_\infty$, say to compute its matrix elements
of the type $\bra{i_1,\ldots,i_n}\rho_\infty\ket{j_1,\ldots,j_n}$ or local observables, is at most {\em polynomial} in $n$.
Since the eigenspaces $\mathfrak{H}^{(\nu)}$ of number-of-holes operator $N_0$ or orthogonal, one can also split decompose the Cholesky factors
$S_n^{(\nu)}(\varepsilon) = \PP^{(\nu)} S_n(\varepsilon)$ 
\begin{equation}
\rho^{(\nu)}_\infty(\varepsilon) = S^{(\nu)}_n(\varepsilon)\,S^{(\nu)\dagger}_n(\varepsilon),
\end{equation}
since $S^{(\nu)}S^{(\nu')\dagger}= 0\; {\rm if}\; \nu\neq\nu'$.
Projected Cholesky factor satisfies a {\em projected} defining relation (\ref{defining_relation})
\begin{equation}
[H,S^{(\nu)}_{n}]=-\ii\varepsilon \left(s^{3}\otimes S^{(\nu)}_{n-1}-S^{(\nu)}_{n-1}\otimes s^{3}\right),
\end{equation}
and can be expressed in terms of a constrained or {\em microcanonical} MPO
\begin{equation}
S^{(\nu)}_n(\varepsilon)  = \sum_{i_1,j_1\ldots i_n,j_n} \delta_{\left(\sum_x \delta_{i_x,2}\right),\nu} \bra{\rm vac}\bb{L}^{i_1 j_1}\cdots \bb{L}^{i_n j_n}\ket{\rm vac} e^{i_1 j_1}\otimes \cdots \otimes e^{i_n j_n}.
\label{mMPO}
\end{equation}
Note that since $[S^{(\nu)},N_0]=0$, the Kronecker-$\delta$ constraint can just as well be replaced by $\delta_{\left(\sum_x \delta_{j_x,2}\right),\nu}$ as only operators 
$e^{i_1 j_1}\otimes \cdots \otimes e^{i_n j_n}$ for which $\sum_x\delta_{i_x,2}=\sum_x\delta_{j_x,2}$ appear in MPO expansion (\ref{MPO}).

We note two limiting cases of our new solution. For zero hole sector $\nu=0$ one obtains exactly the fully polarized boundary driven isotropic ($XXX$) Heisenberg spin-$1/2$ chain and reproduces the solution of Ref. \cite{prl11b} as formulated in \cite{pip13}. 
The other extreme case ($\nu=n$) is the so-called \textit{dark state}, i.e. a pure state  $\rho^{(\nu=n)}_{\infty}=(e^{22})^{\otimes n} = \ket{2,2\ldots 2}\bra{2,2,\ldots 2}$ which
is unaffected by the dissipation, i.e. it simultaneously annihilated by $\LL_{0}$ and $\DD$, $\LL_{0}\rho^{(n)}_{\infty}=\DD\rho^{(n)}_{\infty}=0$.

\subsection{Grand-canonical nonequilibrium steady state ensemble}
\label{grand_canonical}

Any convex mixture of states $\rho_\infty = \sum_\nu c_\nu \rho^{(\nu)}_\infty$, $c_\nu\in\RaR^+$, is a valid NESS density operator as well, which factorizes (\ref{factorization}) with a Cholesky factor $S_n = \sum_\nu \sqrt{c_\nu} S^{(\nu)}_n$.
Microcanonical constraint in (\ref{mMPO}) seems cumbersome as it prevents facilitating transfer matrices for computation of local observables.
There seems to be a particularly attractive option which overcomes this problem. Namely, one may define a {\em grand canonical nonequilbrium steady state} (gcNESS) ensemble by taking a {\em hole chemical potential} $\mu$ with $c_\nu = \exp(\mu \nu)$:
\begin{equation}
\rho_{\infty}(\varepsilon,\mu)=\sum_{\nu=0}^{n}\exp{(\mu \nu)}\,\rho^{(\nu)}_{\infty}(\varepsilon).
\label{grand_canonical_ens}
\end{equation}
Clearly, the addition theorem for exponential function erases the constraint in MPO expansions:
\begin{eqnarray}
S_n(\varepsilon,\mu) = \sum_{i_1,j_1\ldots i_n,j_n} \bra{\rm vac}\bb{L}^{i_1 j_1}(\varepsilon,\mu)\cdots \bb{L}^{i_n j_n}(\varepsilon,\mu)\ket{\rm vac} e^{i_1 j_1}\otimes \cdots \otimes e^{i_n j_n},\\
\rho_{\infty}(\varepsilon,\mu) = \sum_{i_1,j_1\ldots i_n,j_n}\llvac\vmbb{L}^{i_1 j_1}(\varepsilon,\mu)\cdots\vmbb{L}^{i_n j_n}(\varepsilon,\mu)\rrvac e^{i_1 j_1}\otimes \cdots \otimes e^{i_n j_n},
\end{eqnarray}
where the chemical potential only modifies the components of the Lax operators as
\begin{equation}
\bb{L}^{ij}(\varepsilon,\mu) = \exp\left(\frac{\mu}{2}\delta_{i,2}\right)\bb{L}^{ij}(\varepsilon),\quad
\vmbb{L}^{ij}(\varepsilon,\mu) = \exp{\left(\frac{\mu}{2}(\delta_{i,2}+\delta_{j,2})\right)}\vmbb{L}^{ij}(\varepsilon).
\end{equation}
Moreover, introducing a {\em transfer vertex operator}
\begin{equation}
\vmbb{T}(\varepsilon,\mu) = \sum_{i}\vmbb{L}^{i\,i}(\varepsilon,\mu)=\sum_{i,j}\bb{L}^{ij}(\varepsilon,\mu)\otimes \ol{\bb{L}}^{ij}(\varepsilon,\mu),
\label{TVO}
\end{equation}
we define the \textit{nonequilibrium partition function} and express it via the transfer matrix method
\begin{equation}
\mathtt{Z}_{n}(\varepsilon,\mu) = \tr\left(\rho_{\infty}(\varepsilon,\mu)\right) = \llvac \left(\vmbb{T}(\varepsilon,\mu)\right)^n \rrvac.
\label{NPF}
\end{equation}
The hole chemical potential $\mu$ can be connected to the ensemble averaged filling factor (doping) $r$ via logarithmic derivative of the partition function
\begin{equation}
r := \frac{\ave{\nu}}{n} = \frac{\sum_{\nu=0}^n \nu \exp(\nu \mu) \tr\rho^{(\nu)}_\infty}{n\sum_{\nu=0}^n \exp(\nu \mu) \tr\rho^{(\nu)}_\infty} = n^{-1}\partial_{\mu}\log{(\mathtt{Z}_{n}(\varepsilon,\mu))}.
\label{q}
\end{equation}
As usual, we expect the fluctuations $\ave{\nu^2}/n^2 - r^2$ to be thermodynamically small.

We can make a simple assertion about the thermodynamic behavior of $\mathtt{Z}_n$. In the regime, $n\to \infty$, one can write an asymptotic expansion
\begin{equation}
\log \mathtt{Z}_n(\varepsilon,\mu) = \alpha(\varepsilon,\mu) n + \sum_j \beta_j(\varepsilon,\mu) f_j(n) + o(n),
\label{TL}
\end{equation}
where $f_j(n)$ are all possible -- perhaps non-analytic -- {\em super-linear} dependencies satisfying $\lim_{n\to\infty} \frac{n}{f_j(n)} = 0$ (as we shall argue later the most typical being $f(n) = n\log n$), and $o(n)$ is the standard `little-o' notation.
% Insert line below:
Here we have assumed that the chemical potential $\mu$ is an {\em intensive} quantity, i.e., independent of $n$.
According to the definition (\ref{q}), the doping should be confined to the unit interval, $0 \le r \le 1, \forall n$, so the following identities follow in the thermodynamic limit
\begin{equation}
r(\varepsilon,\mu) = \frac{\partial}{\partial\mu} \alpha(\varepsilon,\mu),\qquad
\frac{\partial}{\partial\mu} \beta_j(\varepsilon,\mu) \equiv 0,
\label{TLI}
\end{equation}
i.e., coefficients in front of all super-linear dependencies {\em can not} depend on chemical potential.

\subsection{Computation of local observables}
\label{observables}

Expectation values of (local) observables can be extracted by facilitating {\em auxiliary vertex operators}.
Let $X_{[x,y]}=\one_3^{\otimes (x-1)}\otimes X\otimes \one_3^{\otimes (n-y)}$ be a generic local observable supported on a sublattice between sites $x$ and $y$. Then, a formal expression
\begin{equation}
\expect{X_{[x,y]}}=\mathtt{Z}_{n}^{-1}(\varepsilon,\mu)\;\tr{(X_{[x,y]}\rho_{\infty}(\varepsilon,\mu))},
\end{equation}
can be calculated from the MPO representation of $\rho_{\infty}(\varepsilon,\mu)$ by \textit{tracing out} the physical space $\mathfrak{H}_{\rm s}$ and associating
to each observable $X_{[x,y]}$ a corresponding vertex operator via a mapping $\Lambda_{\ell}:\mathfrak{H}^{\otimes \ell}_{\rm 1}\ra \mathfrak{H}_{\rm a}\otimes \ol{\mathfrak{H}}_{\rm a}$, where $\ell = y-x + 1$,
using the prescription 
\begin{equation}
\Lambda_{\ell}(X)=\vmbb{X}:=\sum_{i_1,j_1\ldots i_\ell,j_\ell} \tr\left((e^{i_1 j_1}\otimes\cdots\otimes e^{i_\ell j_\ell})X\right) \vmbb{L}^{i_1 j_1}\cdots \vmbb{L}^{i_\ell j_\ell}.
\end{equation}
For a complementary part of a lattice, i.e. where $X_{[x,y]}$ operates trivially, one has the transfer vertex operator $\vmbb{T}=\Lambda_{1}(\one_{3})$, eq. \eref{TVO},
so the final expectation value reads
\begin{equation}
\expect{X_{[x,y]}} = \mathtt{Z}_{n}^{-1}\llvac \vmbb{T}^{x-1}\;\vmbb{X}\;\vmbb{T}^{n-y}\rrvac.
\end{equation}
For example, for on-site observables we have auxiliary vertex operators $\Lambda_{1}(e^{ij}) = \vmbb{L}^{j\,i}$, e.g. for magnetization density 
$\Lambda_{1}(s^3) = \vmbb{L}^{11}-\vmbb{L}^{33}$.

As for two point observables, we consider an interesting example of the current density tensor
\begin{eqnarray}
\Lambda_2(J^{ij})=\vmbb{J}^{ij}&=& \ii\left(\vmbb{L}^{j\,i}\vmbb{L}^{ij}\ - \vmbb{L}^{ij}\vmbb{L}^{j\,i}\right)\cr
&=&\ii\sum_{k,l}\left(
\bb{L}^{j\,k}\bb{L}^{i\,l}\otimes \ol{\bb{L}}^{i\,k}\ol{\bb{L}}^{j\,l} - 
\bb{L}^{i\,k}\bb{L}^{j\,l}\otimes \ol{\bb{L}}^{j\,k}\ol{\bb{L}}^{i\,l}
\right).
\label{current_vertex}
\end{eqnarray}
Stationarity (time-independence) of NESS and continuity equation \eref{continuity_equation} imply spatial-independence of current expectation values. In auxiliary transfer matrix formulation \eref{TVO} this implies commutation of transfer vertex operator with current vertex operators when when projected onto subspace of states created upon action of $\vmbb{T}$ on the vacua, namely
\begin{equation}
\bbra{\varphi^{\rm{L}}_{k}}[\vmbb{T},\vmbb{J}^{ij}]\kket{\varphi^{\rm{R}}_{l}}=0,\quad
\bbra{\varphi^{\rm{L}}_{k}}:=\llvac \vmbb{T}^{k},\quad \kket{\varphi^{\rm{R}}_{k}}:=\vmbb{T}^{k}\rrvac.
\end{equation}
Additionally, using representation given in Theorem 2 and highest weight nature of the vacuum, one can with some effort express the expectation values of total current operators \eref{totalJ} in terms of the nonequilibrium partition function (\ref{NPF})
\begin{equation}
\expect{J^{1}}=2\varepsilon\frac{\mathtt{Z}_{n-1}}{\mathtt{Z}_{n}},\quad \expect{J^{3}}=-2\varepsilon\frac{\mathtt{Z}_{n-1}}{\mathtt{Z}_{n}}.
\label{current_recurrence}
\end{equation}
Using parametrization of thermodynamic scaling (\ref{TL}) we can express large $n$ asymptotics of the spin current $J^{\rm s} = J^1 - J^3$ as
\begin{equation}
\log \ave{J^{\rm s}_x} = -\frac{\partial}{\partial n} \log \mathtt{Z}_n + {\rm const} = -\sum_j \beta_j f'_j(n) + {\rm const}.
\end{equation}
For example, in the limiting case $r\to 0$ of $XXX$ spin 1/2 chain, we have \cite{prl11b} a single term in the sum of \eref{TL} with $f_1(n)=n \log n$ and $\beta_1 = 2$, implying a sub-diffusive scaling $\ave{J^{\rm s}_x} \propto n^{-2}$. 
We claim that such scaling may be quite generic, yielding a power-law scaling of the current, $\beta_1$ being the power-law exponent.

In order to obtain more precise, or explicit results on the thermodynamics of observables in our 
nonequilibrium model one would need to have a better understanding of the algebra of auxiliary vertex operators generated by $\vmbb{L}^{ij}$ and of analytic properties of the partition function $\mathtt{Z}_n$, such as in the case of $XXX$ model  \cite{prl11b,iz,pks}. 
A very attractive question would be two investigate $\varepsilon-\mu$ phase diagram of the open Lai--Sutherland chain and to analyze possibilities of {\em nonequilibrium phase transitions}.

For example, one may define the minimal and maximal doping, accessible by an intensive ($n-$independent) chemical potential in the non-equilibrium grand-canonical state, 
as
$r_{\pm}:=\lim_{\mu\to\pm\infty} \lim_{n\to\infty} n^{-1} \partial_\mu \log\mathtt{Z}_n(\varepsilon,\mu)$ (note the importance of the order of the limits!).
Depending on the tails of the $\nu-$dependen\-ce of $\tr\,\rho_\infty^{(\nu)}$, one may have $r_-=0$, or $r_->0$ (and $r_+=1$, or $r_+ < 1$). In the latter of the case(s) 
one may hence expect a phase transition at $r = r_{\pm}$,
whereas the rest of the doping range, $[0,r_-]$ (or $[r_+,1]$), is only accessible by considering a carefully chosen $n-$dependent chemical potential $\mu(n)$. 
Eqs. (\ref{TLI}) imply that the current scaling exponent $\beta_1(r)$ is constant on the entire range $[r_-,r_+]$, nevertheless it may be 
different than the $XXX$ exponent $\beta_1|_{r=0}=2$, which can be obtained from our solution of the Lai-Sutherland chain via different order of the limits $\lim_{n\to\infty}\lim_{\mu\to-\infty}$.
It is thus in principle possible to find even a normal diffusive exponent $\beta_1=1$ and/or transitions to other, say super-diffusive or ballistic behaviors with changing the doping $r$.
Investigating these exciting questions will be a subject of intense future work.

\subsection{The solution as a walking graph state}
\label{walks}

In Ref.~\cite{prl14} a universal interpretation of NESS density operators of integrable boundary driven chains have been given in terms of {\em walking graph states} (WGS). WGS can be considered as an appealing
and compact formulation of matrix product state with infinite dimensional matrices having a simple local structure.

Following notation of \cite{prl14} we show here that our MPO solution (\ref{MPO}) can be given a WGS interpretation as well.
Let the set of vertices of the graph $G$ be an octant of a three-dimensional Cartesian grid 
${\cal V}(G) = \{(j,k,l); j,k,l\in\ZZ^+\}$. The set of edges is a union ${\cal E}(G)={\cal E}'\cup{\cal E}''$ of non-degenerate ${\cal E}'$ and
degenerate ${\cal E}''$ ones. Non-degenerate edges are givens as eight types of pairs of neighboring vertices,
\begin{eqnarray}
{\cal E}' = \{\!\!\!\!\!\!\!\!\!\!\!\!\!\!&((j,k,l),(j+1,k,l)),\; ((j+1,k,l),(j,k,l)),\cr 
                         &((j,k,l),(j,k+1,l)),\; ((j,k+1,l),(j,k,l)),\cr 
                         &((j,k,l),(j+1,k+1,l)),\; ((j+1,k+1,l),(j,k,l)),\cr
                         &((j,k,l),(j,k,l+1)),\; ((j,k,l+1),(j,k,l)); \; j,k,l\in\ZZ^+ \},\!\!\!\!\!\!\!\!\!\!\!\!\!\!\!\!\!\!\!\!\!\!\!\!\!\!\!\!\!\!\!\!
\end{eqnarray} 
corresponding, respectively, to the following values of an {\em index-function} 
$\omega : {\cal E}(G) \to \End(\mathfrak{H}_1)$, namely, $e^{12},e^{21},e^{23},e^{32},e^{13},e^{31},e^{13},e^{31}$.
Edges \begin{equation}
{\cal E}''=\{ ((j,k,l),(j,k,l);i); \; j,k,l\in\ZZ^+, i\in\{1,2,3\} \},
\end{equation} 
are diagonal self-connections and are triple degenerate, corresponding to index function $\omega = e^{i\,i}$.
Finally, we define an amplitude function $a : {\cal E}(G) \to \CC$ by the following prescription.
For each $g \in {\cal E}(G)$ connecting vertex $p(g)$ to vertex $q(g)$ we define $a(g) = \bra{p(g)}\bb{L}^{i(g)j(g)}\ket{q(g)}$, following (\ref{relabeling},\ref{algebra_generators},\ref{bosons},\ref{verma}), where indices $i(g),j(g)$ are determined by the value of index function at $g$, $\omega(g) \equiv e^{i(g)j(g)}$.

Clearly, the MPO \eref{MPO} can now be written as a WGS, i.e., a sum over a set of all walks ${\cal W}_n \ni \ul{g}\equiv(g_1,g_2,\ldots g_n)$ starting at
the origin $(0,0,0)$ and returning to the origin in exactly $n$ steps, $p(g_1) = (0,0,0), q(g_x) = p(g_{x+1}), q(g_n) = (0,0,0)$,
\begin{equation}
S_n = \!\!\!\sum_{(g_1,g_2\ldots g_n)\in{\cal W}_n}\!\!\!\! a(g_1)a(g_2)\cdots a(g_n)\, \omega(g_1)\otimes \omega(g_2)\otimes \cdots\otimes \omega(g_n).
\end{equation}
Contrary to $XXZ$ and Hubbard models \cite{prl14}, where each value $\omega(g)$ of the index function corresponds to {\em only one} direction $q(g)-p(g)$ in the graph diagram, and consequently the
partition function could be written as an appealing walking graph sum of strictly positive terms
$\mathtt{Z}_n = \sum_{(g_1,g_2\ldots g_n)\in{\cal W}_n} |a(g_1) a(g_2) \cdots a(g_n)|^2
$, (even if $a(g)$ are not $\CC$-numbers like in the Hubbard case) this is {\em not} the case here,
since the index function is {\em degenerate}. E.g., to $e^{13}$ there correspond directions $(1,1,0)$ and $(0,0,1)$.
Nevertheless, one can verify that the whole partition function still remains a sum of positive terms being attributed to multiple
walks $\ul{g}$ which share common index functions $\omega(\ul{g}):=\omega(g_{1})\otimes \cdots \otimes \omega(g_{n})$,
i.e., now individual contributions from degenerate walks coherently add up to a final amplitude, much like the interference property in standard wave-like phenomena.

\subsection{Characterization of the Lie algebra.}

\label{lie_algebra}

The Lie algebra $\mathfrak{g}$, eq. (\ref{algebra}), has a non-trivial structure.
It can be decomposed however (according to Levi theorem) as a semi-direct product of a solvable ideal (radical) and semi-simple part,
\begin{equation}
\mathfrak{g}=\mathfrak{r}\ltimes \mathfrak{a}.
\label{Levi_decomposition}
\end{equation}
In our case $\mathfrak{a}$ is given by $\rm{lsp}\{\bb{v}^+,\bb{v}^-,\bb{l}^{+}\}$, writing $\bb{l}^\pm:=\bb{l}^{\ua}\pm\bb{l}^{\da}$, i.e. $\mathfrak{a}\cong \mathfrak{sl}_{2}$ is
isomorphic to spin algebra, whereas $\mathfrak{r}=\rm{lsp}\{\bb{t}^{\pm},\bb{u}^{\pm},\bb{l}^-,\bb{l}^{0}\}$, generates a (non-nilpotent) radical.
The element $\bb{l}^{0}$ lies in the center of $\mathfrak{g}$.
It is worth noticing also, that parameter $\eta$ can be fully removed from the algebra (\ref{algebra}) by diving all generators by $\eta$, except $\bb{t}^+, \bb{u}^-$.

\subsection{Symmetries of the Lax and transfer operators}

\label{symmetry_lax}

In contrast to situation with $XXX$ or $XXZ$ spin 1/2 chain \cite{prl11b,kps}, it might seem surprising here that the fundamental local unit, the Lax operator $\bb{L}$,
does not exhibit the full {\em SL(3)} symmetry of the interaction.
However, the dissipative driving breaks the {\em SL(3)} symmetry, resulting in only remaining $U(1)$ global symmetry of the Liouvillian flow
generated by 
\begin{equation} 
M = \sum_{x=1}^n s^3_x,
\label{M}
\end{equation}  
over $\End(\mathfrak{H}_{\rm s})$.
Consequently, we found generators of {\em U(1)} symmetry for the Lax operators represented in  $\mathfrak{H}_{\rm 1}\otimes \mathfrak{H}_{\rm a}$, and
$\mathfrak{H}_{\rm 1}\otimes \mathfrak{H}_{\rm a}\otimes \ol{\mathfrak{H}}_{\rm a}$, namely
\begin{eqnarray}
&[\bb{L},\ii \varepsilon s^{3}\otimes \one_{\rm a} + \one_3 \otimes \bb{l}^{+}]=0, \\
&[\vmbb{L},\ii \varepsilon s^{3}\otimes \one_{\rm a}\otimes\one_{\rm a} + \one_3\otimes \bb{l}^{+} \otimes \one_{\rm a} + \one_3 \otimes \one_{\rm a}\otimes \ol{\bb{l}}^{+}] = 0.
\label{U1_symmetry}
\end{eqnarray}
It remains to be investigated whether other gauges exist in which Lax operators exhibit non-Abelian symmetry.

Furthermore, transfer vertex operator $\vmbb{T}$ exhibits $U(1)\times U(1)$ symmetry, i.e. there exist two conserved auxiliary-space operators $\vmbb{K}^{\pm} \in \End(\mathfrak{H}_{\rm a} \otimes \ol{\mathfrak{H}}_{\rm a})$,
\begin{equation}
[\vmbb{T},\vmbb{K}^\pm] = 0,\qquad \vmbb{K}^\pm := \bb{l}^\pm \otimes\one_{\rm a} + \one_{\rm a} \otimes  \ol{\bb{l}}^\pm.
\end{equation}
Conserved operators look particularly useful in the auxiliary spin-boson representation of Theorem 2 (note that $\ol{\eta}=-\eta$ for physical (real) dissipation) where we have now four independent bosonic modes $\bb{b}_{\uparrow \downarrow},\ol{\bb{b}}_{\uparrow \downarrow}$ and two
complex spins $\bb{s}^\alpha$ and $\ol{\bb{s}}^\alpha$ with representation parameters $p=1/2-1/\eta$ and $\ol{p}=1/2+1/\eta$
\begin{eqnarray}
\vmbb{K}^+ &=& \eta \left(\bb{b}^\dagger_\uparrow \bb{b}_\uparrow + \bb{b}^\dagger_\downarrow \bb{b}_\downarrow - 2 \bb{s}^\z\right) -  \eta \left(\ol{\bb{b}}^\dagger_\uparrow \ol{\bb{b}}_\uparrow + \ol{\bb{b}}^\dagger_\downarrow \ol{\bb{b}}_\downarrow - 2 \ol{\bb{s}}^\z\right),\\
\vmbb{K}^- &=& \eta \left(\bb{b}^\dagger_\uparrow \bb{b}_\uparrow - \bb{b}^\dagger_\downarrow \bb{b}_\downarrow\right) -  \eta \left(\ol{\bb{b}}^\dagger_\uparrow \ol{\bb{b}}_\uparrow - \ol{\bb{b}}^\dagger_\downarrow \ol{\bb{b}}_\downarrow\right).
\end{eqnarray}
Computation of nonequilibrium partition function (\ref{NPF}) should hence be performed on a 4D sub-lattice of a 6D lattice (a basis of $\mathfrak{H}_{\rm a}\otimes \ol{\mathfrak{H}}_{\rm a}$,
$\{ \ket{j,k,l,\ol{j},\ol{k},\ol{l}}; j,k,l,\ol{j},\ol{k},\ol{l}\in\ZZ^+\}$) where, say $\ol{j},\ol{k}$ can be eliminated using constraints:
\begin{equation}
j-k=\ol{j}-\ol{k},\quad j + k - 2l = \ol{j} +\ol{k} - 2\ol{l}.
\end{equation}
This is analogous to `diagonal reduction' of the transfer matrix for the open $XXZ$ chain proposed in Refs.~\cite{prl11b,prl13}.

\subsection{Symmetries of the Liouvillian flow and its fixed point}
\label{symmetries}

Besides the strong {\em U(1)} symmetry generated by $N_0$, the full Liouvillian flow has another {\em U(1)} symmetry generated by magnetization operator $M$ \eref{M}, as noted in subsect. \ref{symmetry_lax}.
This is a {\em weak} symmetry in the sense of Ref.~\cite{buca} and can be formally written as
\begin{equation}
[M,\LL\rho] = \LL([M,\rho]), \quad \forall \rho.
\end{equation}
As a consequence $M$ should be a `good quantum number' for the fixed point (NESS) $\rho_\infty$, i.e., $[\rho_\infty,M]=0$ and 
\begin{equation}
\bra{i_1,\ldots,i_n}\rho_\infty\ket{j_1,\ldots,j_n}\neq 0\quad {\rm only\; if}\quad \sum_{x=1}^n i_x = \sum_{x=1}^n j_x.
\end{equation}

The Liouvillian flow and NESS display additional $\ZZ_2$-parity symmetry which is a composition of
\textit{lattice reversal} $\hat\mathcal{R}\in \End{(\mathfrak{F})}$,
\begin{equation}
\hat\mathcal{R}(e^{i_1 j_1}\otimes e^{i_2 j_2}\otimes\cdots\otimes e^{i_n j_n}) = e^{i_n j_n}\otimes e^{i_{n-1} j_{n-1}}\otimes\cdots\otimes e^{i_1 j_1},
\label{lattice_reversal}
\end{equation}
and a product of local mirror symmetries $\hat{\mathcal{S}}\in \End{(\mathfrak{F})}$ which exchange spin-up and spin-down particles,
\begin{equation}
\hat\mathcal{S} = \hat\mathcal{S}_1^{\otimes n},\quad
\hat\mathcal{S}_1(e^{ij}) = e^{3-i+1,3-j+1},
\label{skew_flip}
\end{equation}
namely 
\begin{equation}
[\hat\mathcal{R}\,\hat\mathcal{S},\LL] = 0 \qquad {\rm and} \qquad
\hat\mathcal{R}\,\hat\mathcal{S}\,\rho_\infty = \rho_\infty.
\end{equation}

Cholesky factor $S_{n}(\epsilon)$ however acquires another $\ZZ_2$-parity symmetry. By means of \textit{transposition map}
$\hat\mathcal{T}\in \End{(\mathfrak{F})}$,
\begin{equation}
\hat\mathcal{T} = \hat\mathcal{T}_1^{\otimes n},\quad \hat\mathcal{T}_1(e^{ij}) = e^{j\,i},
\label{transposition_map}
\end{equation}
one finds
\begin{equation}
\hat\mathcal{R}\,\hat\mathcal{S}\, S_{n}=\hat\mathcal{T}\hat\mathcal{S}\,S_{n}=S_{n}.
\label{S_parities}
\end{equation}
Notice that $\hat\mathcal{T}\hat{\mathcal{S}}$ pertains to the symmetry with respect to exchange of the bosonic modes in auxiliary space.

\subsection{Transfer matrix property of Cholesky factors}
\label{transfer_matrix}

Similarly to $XXX$ \cite{pip13} and Hubbard \cite{prl14} chains the Cholesky factor is found, empirically by checking explicitly systems of small size $n$, to exhibit a transfer matrix properly, namely
\begin{equation}
[S_n(\varepsilon),S_n(\varepsilon')] = 0,\quad \forall \varepsilon,\varepsilon' \in \CC.
\label{TMP}
\end{equation}
This property justifies calling the $\bb{L}$-operator a \textit{quantum Lax matrix} with $\bb{M}$-opera\-tor being a corresponding monodromy matrix and $S_n(\varepsilon) = \lvac \bb{M}(\varepsilon)\rvac$ the corresponding
transfer matrix where the trace in infinitely dimensional auxiliary space is replaced by a ground state expectation value \cite{pip13}.
It remains an open issue though to prove that $\bb{L}(\varepsilon)$ belongs to solutions of the \textit{quantum Yang-Baxter equation}.
Clearly, the property (\ref{TMP}) can be extended to grand-canonical objects due to orthogonality of subspaces $\mathfrak{H}^{(\nu)}$, namely
$[S_n(\varepsilon,\mu),S_n(\varepsilon',\mu')]=0$.

\section{Conclusions}
\label{conclusions}

We have presented an explicit infinite rank matrix-product state construction of an exact solution for a grand-canonical nonequilibrium steady state of
boundary-driven integrable {\em SU(3)}-symmetric spin-$1$ chain. Beside the external chemical potential, controlling an average filling factor of
conserved ``hole particles'', the NESS (continuously) depends on also on the bath coupling parameter,
describing strength of incoherent processes at the boundaries. Quite remarkably, the elements of the main building block (the
$\bb{L}$-operator) generate a Lie algebra of non-trivial structure whose simple part is given by classical $\mathfrak{sl}_{2}$ algebra.
Despite the fact that $\bb{L}$-operator does not exhibit invariance with respect to any continuous non-Abelian symmetry, empiric evidence clearly suggests
that it generates a quantum transfer matrix of an (abstract) integrable system,
indicating that a Yang-Baxter structure is sitting underneath.
Another central aspect to the problem is that the auxiliary space can be factored into three-fold product of \textit{infinite-dimensional}
quantum spaces -- a Fock space of two independent bosonic modes and one generic  representation of $\mathfrak{sl}_{2}$ (Verma module) -- depending on
one complex continuous representation (spin) parameter $p$.
In order to fulfill the boundary system of equations which guarantee solutions to our problem, the value of spin parameter must be chosen according
with the dissipation rate. The solution contains, as a special extreme case, previously known NESS for symmetrical driving of the spin-$1/2$ (isotropic) Heisenberg model.
It remains an open issue whether presented solution admits any integrable continuous deformation ($q-$deformation), enabling generalization to anisotropic versions of the Lai--Sutherland model, say
for the Perk-Schultz model \cite{perk_schultz}. Another interesting open issue is generalizing our solution to more than three ($N > 3$) component Sutherland models.

\section*{Acknowledgements}
E.I. acknowledges discussion with \v{S}. \v{S}penko. The work has been supported by grants P1-0044 and J1-5439 of Slovenian Research Agency.


\section*{References}
 
\begin{thebibliography}{10}

\bibitem{spinchains} 
A. V. Sologubenko, T. Lorenz, H. R. Ott and A. Freimuth, J. Low Temp. Phys. {\bf 147}, 387 (2007)

\bibitem{fabian}  F. Heidrich-Meisner, A. Honecker, and W. Brenig, Eur. Phys. J. Special Topics {\bf 151}, 135 (2007).  

\bibitem{affleck}  J.~Sirker, R.~G.~Pereira and I.~Affleck, Phys. Rev. Lett. {\bf 103}, 216602 (2009); Phys. Rev. B {\bf 83}, 035115 (2011).

\bibitem{marko} M. \v Znidari\v c, Phys. Rev. Lett. {\bf 106}, 220601 (2011). 

\bibitem{gemmer} J.~Gemmer, M.~Michel and G.~Mahler, {\em Quantum Thermodynamics}, 2nd Ed. (Springer-Verlag, Heidelberg, 2009).

\bibitem{prl108} T. Prosen and I. Pi\v zorn, Phys. Rev. Lett. {\bf 101},  105701 (2008).

\bibitem{zoller} A.~Tomadin, S.~Diehl and P.~Zoller, Phys. Rev. A {\bf 83},  013611 (2011).

\bibitem{gemmer2}  H. Wichterich, M.-J. Henrich, H.-P. Breuer, J. Gemmer, and M. Michel, Phys. Rev. E {\bf 76}, 031115 (2007).

\bibitem{derrida} B.~Derrida, M.~R.~Evans, V.~Hakim and V.~Pasquier, J. Phys. A: Math. and Gen. {\bf 26}, 1493 (1993).

\bibitem{schutz} G. M. Sch\" utz, {\em Exactly Solvable Models for Many-Body Systems Far from Equilibrium}, in ``Phase transitions and Critical Phenomena'', Vol. 19, Eds.
C. Domb and J. L. Lebowitz, (Academic Press, London 2001).

\bibitem{blythe} R.~A.~Blythe and M.~R.~Evans, J. Phys. A: Math. and Theor. {\bf 40}, R333 (2007).

\bibitem{njp} T. Prosen, New J. Phys, {\bf 10}, 043026 (2008).

\bibitem{prl11a} T. Prosen, Phys. Rev. Lett. {\bf 106}, 217206 (2011).

\bibitem{prl11b} T. Prosen, Phys. Rev. Lett. {\bf 107}, 137201 (2011).

\bibitem{kps} D.~Karevski, V.~Popkov and G.~M.~Sch\" utz, Phys. Rev. Lett. {\bf 110}, 047201 (2013).

\bibitem{prl14} T. Prosen, Phys. Rev. Lett. {\bf 112}, 030603 (2014).

\bibitem{pip13} T. Prosen, E. Ilievski and V. Popkov, New J. Phys. {\bf  15}, 073051 (2013).

\bibitem{iz} E.~Ilievski and B.~\v Zunkovi\v c, J. Stat. Mech., P01001 (2014).

\bibitem{prl13} T. Prosen and E. Ilievski, Phys. Rev. Lett. {\bf 111}, 057203 (2013).

\bibitem{sutherland_book} B.~Sutherland, {\em Beautiful Models}, (World Scientific, Singapore 2004).

\bibitem{korepin} V.~E.~Korepin, N.~M.~Bogoliubov and A.~G.~Izergin, {\em Quantum Inverse Scattering Method and Correlation Functions}, (Cambridge University Press, Cambridge, 1993).

\bibitem{derkachev} S.~Derkachov, D.~Karakhanyan and R.~Kirschner,  Nucl. Phys. B {\bf 618}, 589 (2001).

\bibitem{lai} J.~K.~Lai, J. Math. Phys. {\bf 15}, 1675 (1974).

\bibitem{sutherland} B. Sutherland, Phys. Rev. B {\bf 12}, 3795 (1975).

\bibitem{uimin} G.~Uimin, JETP Lett. {\bf 12}, 225 (1970).

\bibitem{tJ} J.~Ambjorn, D.~Karakhanyan, M.~Mirumyan, A.~Sedrakyan, Nucl. Phys. {\bf B 599}, 547 (2001).

\bibitem{lindblad} G. Lindblad, Comm. Math. Phys. {\bf 48}, 119 (1976).

\bibitem{gorini} V. Gorini, A. Kossakowski, and E. C. G. Sudarshan, J. Math. Phys. {\bf 17}, 821 (1976).

 \bibitem{buca} B.~Bu\v ca and T.~Prosen, New. J. Phys. {\bf 14}, 073007 (2012).

\bibitem{albert} V.~V.~Albert and L.~Jiang, {\tt arXiv:1310.1523}.

\bibitem{evans} D.~E.~Evans, Comm. Math. Phys. {\bf 54}, 293 (1977).

\bibitem{pks} V.~Popkov, D.~Karevski and G.~M.~Sch\" utz, Phys. Rev. E {\bf 88}, 062118 (2013).
 
\bibitem{perk_schultz} C.~L.~Schultz, Phys. Rev. Lett. {\bf 46}, 629 (1981); J.~H.~H.~Perk and C. L. Schultz, Phys. Lett. {\bf 84A}, 407 (1981).

\end{thebibliography}
\end{document}